\documentclass{article}

\usepackage{PRIMEarxiv}

\usepackage[utf8]{inputenc} 
\usepackage[T1]{fontenc}    
\usepackage{hyperref}       
\usepackage{url}            
\usepackage{booktabs}       
\usepackage{amsfonts}       
\usepackage{nicefrac}       
\usepackage{microtype}      
\usepackage{lipsum}
\usepackage{fancyhdr}       
\usepackage{graphicx}       
\graphicspath{{media/}}     

\usepackage[utf8]{inputenc}     
\usepackage{amssymb,amsmath}    
\usepackage{url}
\usepackage{graphicx}           
\usepackage{xcolor}             
\usepackage{xspace}             
\usepackage{hyperref}           

\newcommand{\ie}{\emph{i.e.}\xspace}
\newcommand{\eg}{\emph{e.g.}\xspace}
\newcommand{\et}{\emph{et al.}\xspace}

\usepackage{mathrsfs}

\usepackage{booktabs} 

\usepackage{multirow} 

\title{Super-Linear Growth and Rising Inequality in Online Social Communities: Insights from Reddit
}

\author{
  Guilherme Machado \\
  University of Aveiro \\
   \And
  Diogo Pacheco \\
  University of Exeter \\
   \And
  Ronaldo Menezes \\
  University of Exeter \\
   \And
  Gareth Baxter \\
  University of Aveiro \\
}

\begin{document}
\maketitle

\begin{abstract}
We study the effect of the number of users on the activity of communities within the online content sharing and discussion platform Reddit, called subreddits. We found that comment activity on Reddit has a heavy-tailed distribution, where a large fraction of the comments are made by a small set of users. Furthermore, as subreddits grow in size, this behavior becomes stronger, with activity (measured by the comments made in a subreddit) becoming even more centralised in a (relatively) smaller core of users. We verify that these changes are not explained by finite size nor by sampling effects. Instead, we observe a systematic change of the distribution with subreddit size. To quantify the centralisation and inequality of activity in a subreddit, we used the Gini coefficient. We found that as subreddits grow in users, so does the Gini coefficient, seemingly as a natural effect of the scaling. 
We found that the excess number of comments (the total number of comments minus the total number of users) follows a power law with exponent 1.27. 
For each subreddit we considered a snapshot of one month of data, as a compromise between statistical relevance and change in the system’s dynamics. We show results over the whole year 2021 (with each subreddit having twelve snapshots, at most), nevertheless all results were consistent when using a single month or different years.
\end{abstract}


\section{Introduction}
Social media have become an integral part of individuals' daily lives, serving as a primary means of social interaction, news dissemination, and information sharing~\cite{whiting2013people,gil2012social}. Given that online interactions are a relatively recent phenomenon, humans have not biologically evolved to engage predominantly in virtual communication. In turn, the nature of interactions on social media platforms tends to mirror those found in the physical world, incorporating familiar social behaviours and structures~\cite{hazarie2020uncovering,schnauber2023routines}. However, unlike face-to-face interactions, virtual communication is not constrained by spatial limitations, thereby facilitating a higher volume and broader scope of interactions. While it is well-established that inequalities arise in real-world social structures due to various factors~\cite{lin2000inequality,piketty2014inequality,collins2015intersectionality}, the development and manifestation of inequalities within online interactions remain insufficiently explored.

Previous studies have explored the statistical properties of social interactions across various platforms, consistently identifying broad, heavy-tailed distributions such as power laws in the dynamics of user engagement \cite{myers2014bursty,muchnik2013origins}. These findings indicate that user interactions often follow predictable scaling laws, akin to those observed in natural and social phenomena, although, interestingly, some research has found that not all social media interactions follow pure power laws~\cite{kwak2010what, vranic2022universal}. Furthermore, it is well acknowledged that many social media platforms introduce inherent biases through algorithms that govern content visibility, interaction facilitation, and group formation~\cite{bozdag2013bias,stoica2018algorithmic}. These algorithmic mechanisms can distort natural interaction patterns, making it challenging to determine whether observed statistical regularities are intrinsic to human behaviour or are a result of platform-specific interventions driven by commercial interests \cite{helmond2015platformization,cotter2019playing}. Consequently, to develop more generalisable hypotheses regarding online social interactions, it is imperative to examine platforms with minimal algorithmic interference in content curation and interaction dynamics. While achieving complete impartiality is nearly impossible, Reddit~\cite{redditinc} stands out as a promising candidate due to its comparatively open platform structure, which allows for more organic interaction dynamics and the formation of user-driven communities. This openness reduces the extent of algorithm-induced biases, thereby providing a more authentic environment to study user activity and interaction patterns without the pervasive influence of commercial algorithmic manipulation \cite{fu2024unravelling}. As such, Reddit serves as a better setting for investigating the fundamental principles governing online social interactions and the emergence of inequalities within these virtual communities~\cite{medvedev2019modelling}.

While Reddit differs from other social media in some key respects, it nevertheless exhibits many of the same statistical features as other platforms \cite{medvedev2019anatomy}. Reddit distinguishes itself through its organisation around topical communities known as {\em subreddits}. Unlike platforms such as X 
(formerly Twitter), Facebook, or Instagram, which are primarily centred on individual profiles as content producers, Reddit's structure allows users to engage within diverse, interest-specific communities. Users can subscribe to multiple subreddits, each dedicated to a particular subject or theme, ranging from broad topics like news and technology to niche interests such as specific hobbies or regional communities. The size of these subreddits varies widely, hosting anywhere from a handful of active users (\eg, \textit{r/londonfootballmeetup}, \textit{r/learnAI}, \textit{r/starwarslego}) to hundreds of thousands of participants (\eg, \textit{r/AskReddit}, \textit{r/memes}, \textit{r/wallstreetbets}). This community-centric model fosters specialised discussions and enables the formation of tightly-knit groups within the larger Reddit ecosystem.

Within subreddits, users have access to a wide array of topics and can engage actively by commenting on posts and other users' contributions. Moreover, Reddit emphasises anonymity and pseudonymity, as users are not required to disclose personal information beyond what is necessary for account creation. This encourages open and honest discourse, although it also presents unique challenges in content moderation and community management. 
Each subreddit is typically moderated by a team of volunteer moderators who enforce community-specific rules, maintain discourse quality, and manage content visibility, such as, by removing or approving posts and comments that violate guidelines, as well as categorising content as `spoilers' or NSFW (not safe for work) when necessary. 
Reddit facilitates community-driven feedback through an upvoting and downvoting system, which plays a pivotal role in determining the visibility and prominence of posts. Additionally, Reddit features a `karma' system~\cite{redditKarma} that reflects a user's reputation within the platform, earned through the reception of upvotes on their posts and comments. In certain subreddits, a minimum karma threshold is even enforced for submitting posts, ensuring a baseline quality of content within the community, which could be seen as algorithm control, but it is more like reputation-based rules. 
These elements collectively contribute to Reddit's distinctive environment, balancing open user participation with structured community governance.
Another notable aspect of Reddit is that the majority of its content—including posts, comments, usernames, profiles, karma scores, upvote/downvote ratios, and associated metadata—is publicly accessible, reflecting its commitment to an `open internet'~\cite{reddit_privacy_policy_info}. This extensive dataset has been systematically collected and made available by the Pushshift project~\cite{Pushshift, baumgartner2020pushshift}.

Dynamic interaction systems frequently result in the dominance of specific groups or individuals, exhibiting patterns reminiscent of the Pareto Principle, where a small proportion of participants account for the majority of contributions or influence within a community. This phenomenon of inequality is observable across various social, economic, and biological systems, underscoring tendencies towards hierarchical structures and power asymmetries. 
We aim to investigate the extent to which online systems exhibit such inequalities and to explore whether there are inherent limits to these disparities. By analysing the distribution of user engagement and activity, we seek to 
contribute
to a more comprehensive understanding of the dominance of certain user groups and equitable participation within digital ecosystems.

In this study, we investigate the statistical properties of user activity on Reddit, with a particular focus on how the number of users within a subreddit influences overall engagement. To quantify user interactions, we measure the number of comments each user contributes over a fixed time window of one month. Our analysis confirms that overall user activity is highly heterogeneous, exhibiting distributions consistent with log-normal or power-law behaviours, as previously observed in~\cite{vranic2022universal}. When examining individual subreddits, we observe that total activity scales super-linearly with subreddit size, as determined by the number of active commenting users. Specifically, we identify a well defined 
power law growth in excess activity, characterised by a growth exponent of approximately 1.3 which is, perhaps surprisingly, in line with the scaling of activity found in social systems in the physical setting of cities, as observed by Bettencourt \et~\cite{bettencourt2010urban}. 
Furthermore, the distribution of comments within subreddits of similar sizes reveals heavy-tailed characteristics, with the shape of these distributions evolving as subreddit size increases. Importantly, we verify that these observed changes are not attributable to finite size or sampling effects across communities of varying sizes. Instead, there is a systematic broadening of the comment distribution correlated with subreddit size. Consequently, the super-linear growth in activity is not uniformly distributed among users; rather, larger subreddits exhibit increasing inequality, with a substantial portion of activity concentrated among a diminishing fraction of users. This inequality is quantified using the Gini coefficient, which we demonstrate increases monotonically with subreddit size. However, this increase slows as subreddits reach substantial sizes, in an apparent plateau, suggesting the existence of inherent limits to such disparities in very large communities.

\section{Data and Methods}

A comprehensive collection of Reddit comments was obtained from the Pushshift project \cite{Pushshift}. The dataset encompasses all comments made on Reddit, organised on a monthly basis according to their time of submission. For each comment, the dataset includes the author’s username, the comment content, the timestamp (UT) of publication, a controversy score, the number of upvotes and downvotes, the author’s karma score, and the UT corresponding to the data scraping moment. The dataset utilised in this study spans from December 2005 to December 2021.

We opted to analyse the data on a monthly basis for each subreddit. This timeframe provides sufficient statistical relevance to capture a `snapshot' of each subreddit's activity without being excessively long, which could obscure changes in the dynamics or structure of Reddit. Additionally, this approach facilitates clear comparisons across different months and years. Consequently, in annual plots, a subreddit that was active throughout the year will appear twelve times, corresponding to each monthly interval. For each month, we filtered the data to construct a specialised database containing the following information: the names of subreddits with at least one comment; anonymised usernames of the users who commented within each subreddit; and the number of comments each user made in the respective subreddit. This approach ensures that our analysis captures active participation without compromising user anonymity. We did not consider the data from user profiles pages, which are defined has "personal subreddits" (starting with "u/", instead of "r/"). The reason being that profiles are not communities like subreddits since they are centred around a profile, and not a topic. Also, the rules of interaction in the profile pages are different, for example, only the owner of the profile is allowed to publish in his own page, while other users may only comment or give up/down votes.

The analysis presented herein focuses on data from the year 2021. However, we observed that the results are both qualitatively and quantitatively consistent across other years (see Supplementary Materials). We selected 2021 as our primary dataset because it exhibited the least statistical fluctuations, attributable to having the highest number of users and the most substantial volume of content generated.

Our study exclusively concentrates on user comments rather than posts or publications. This decision was made because comments are significantly more abundant on Reddit, providing a more robust basis for statistical analysis. 
Additionally, incorporating both comments and posts could introduce complexities due to their potentially distinct dynamics. Future work may extend this research to include posts, allowing for comparative analyses between different types of user-generated content; comments are seen as responses and reactions to posts.

\section{Results}

\subsection{General Statistics of Reddit}

This section presents general statistical features of user activity across the Reddit platform. The goal here is to establish first the general characteristics of the platform before delving into specific analyses around the effect of the size of the subreddits and how the inequality of these groups varies across the sizes. For each month analysed, our work encompassed on average around 200,000 subreddits, comprising comments from approximately 10,000,000 users.
A substantial number of subreddits have only one active user (around 25\%), while the largest subreddit includes well over 1,000,000 active users and millions of comments. 

\begin{figure}[htb]
\centering
\includegraphics[width=0.45\textwidth]{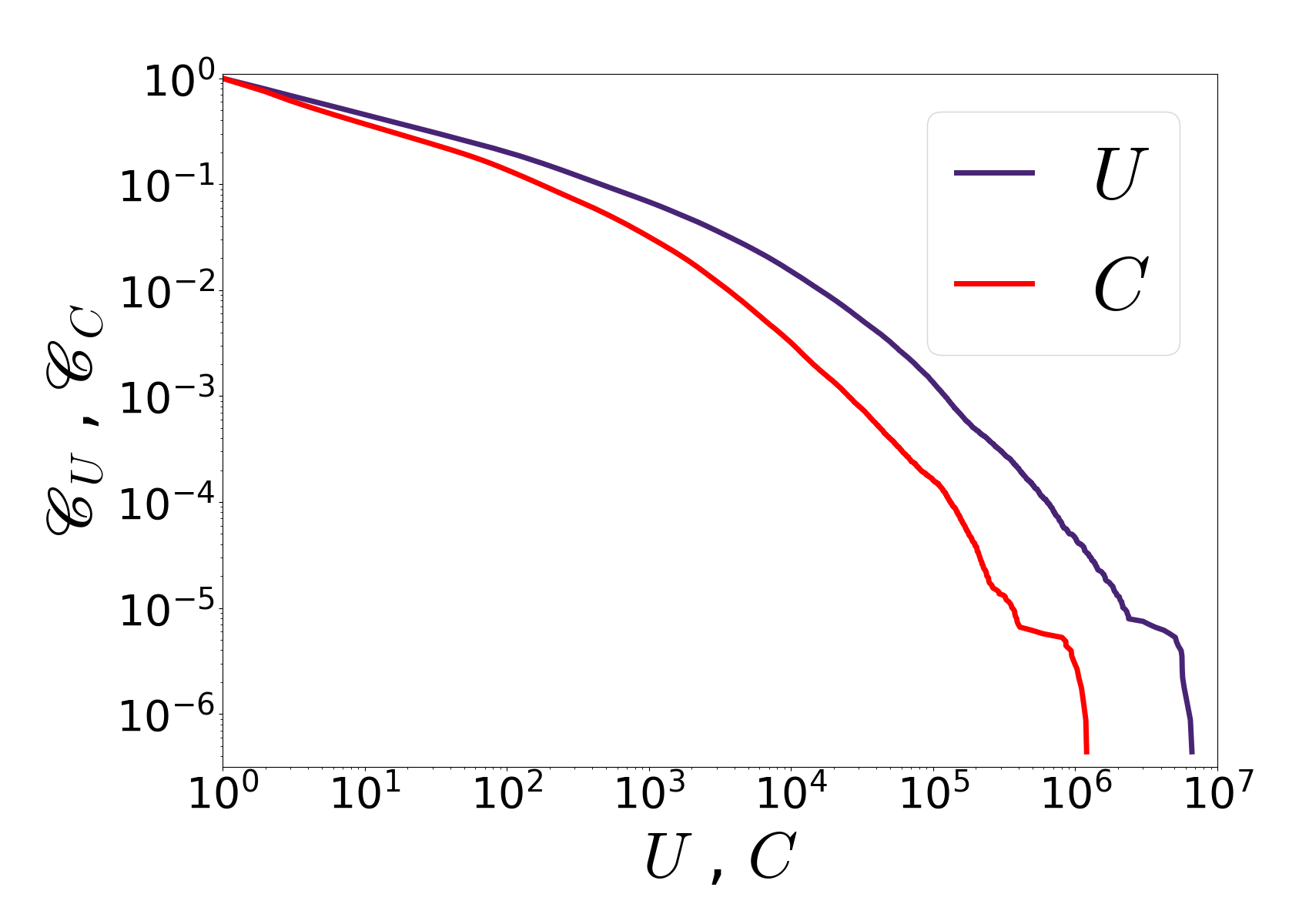}
\includegraphics[width=0.45\textwidth]{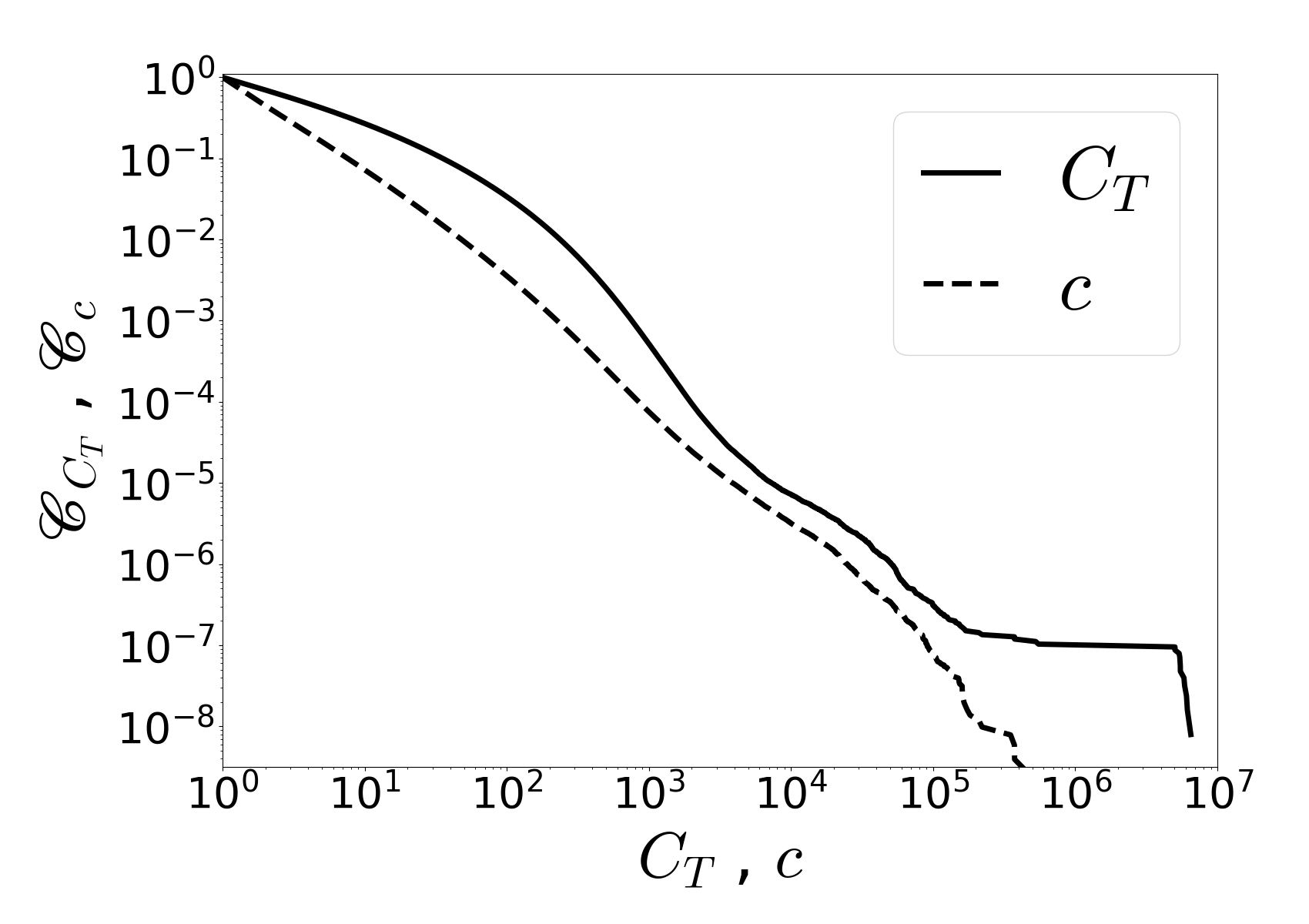}
\caption{(Left Panel) The complementary cumulative probability distribution of the total number of active users per subreddit (upper, purple line), \(U\), and total number of comments made in a subreddit, \(C\), (lower, red line). (Right Panel) The total number of comments made by each user in all subreddits (black solid line), \(C_{T}\), and the general distribution of the comments-per-user in a subreddit, \(c\) (right panel).  Note the log-log scale used to make the very heterogeneous distributions more visible.}
\label{Img_Hist_User_Com}
\end{figure}

We employed two primary metrics to quantify activity within each subreddit: the number of active users in a subreddit during one month, \(U\), (\ie, users who made at least one comment) and the total number of comments made within the subreddit during the same period, \(C\). As illustrated in Figure \ref{Img_Hist_User_Com}, the distributions of both users and comments per subreddit are exceptionally broad, reflecting a vast diversity of activity levels across subreddits. 
These distributions are consistent with log-normal and power-law distributions, aligning with previous findings \cite{vranic2022universal}. We quantified them by using identical numerical methods, and following the approach suggested in \cite{alstott2014powerlaw} for heavy-tailed distributions, as we show in the Supplementary Material.

Additionally, we measured the distribution of individual user activity across Reddit by quantifying the total number of comments each user made within the one-month window, \(C_{T}\). We find that the distribution of comments made by users also exhibits heavy-tailed behaviour, see Figure~\ref{Img_Hist_User_Com}. Although the tail resembles a power-law distribution, the overall distribution is evidently more complex. These results reveal that not only do subreddits vary greatly in activity, but individual users also display significantly different patterns of engagement. In fact, a large fraction of comments is generated by a small fraction of users: we identified a `Pareto principle' where 20\% of users contribute 84\% of the comments; and approximately 50\% of users make only one comment in a whole month on Reddit.

The `global' distribution of comments-per-user in a subreddit, \(C_{c}\) (black dashed line in Figure~\ref{Img_Hist_User_Com}), was calculated by grouping together all the monthly comment-counts of each user in each different subreddit. Interestingly, it closely follows a power law distribution of exponent 
\(-1.44\), and a \(r^2 = 0.99\)
(using a least-squares method for the linear regression). The variation between the distributions of \(c\) and \(C_{T}\) underscores the substantial diversity in engagement, contrasting users total activity with their comments allocation to subreddits.

\section{Variations with Subreddit Size}

In the previous sections, we established that both the total number of comments in a subreddit and the distribution of individual user activity are highly heterogeneous. This section investigates how subreddit size, defined by the number of active users, influences these activity patterns.

\subsection{Total Comments vs. Number of Users}

\begin{figure}[ht] 
\centering
\includegraphics[width=0.7\textwidth]{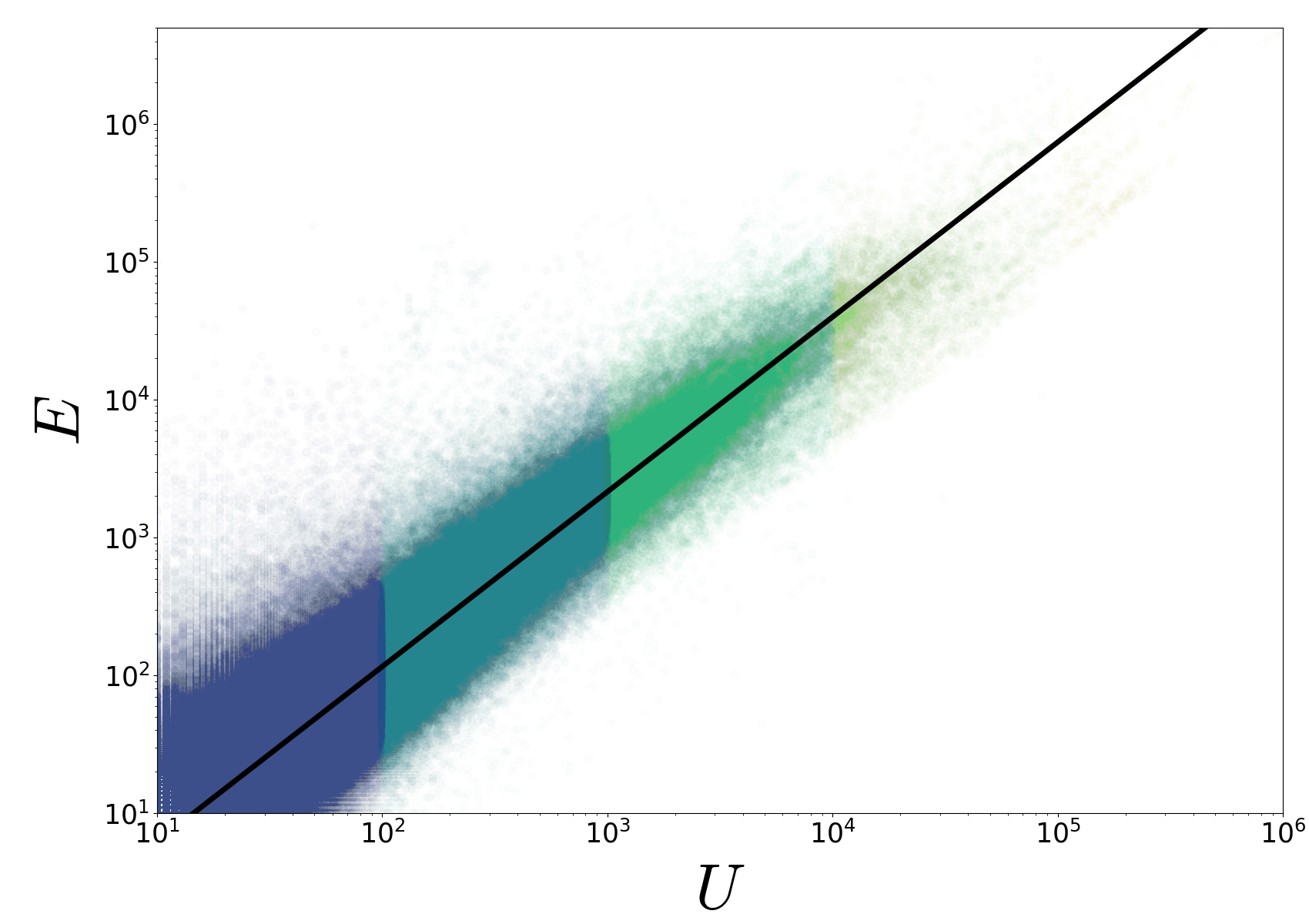}
\caption{The relationship of the excess number of comments, ($E = C-U$), to the number of users, $U$, in each subreddit. Each dot represents a single subreddit, and was attributed a colour depending on its size, $U$, for better relation with later figures. The black line represents the linear regression. The regression only considered subreddits with more than 10 users, due to the discrete nature of the data.}
\label{Img_NCom_vs_Nuser}
\end{figure}

A natural first step is to quantify how the total activity in a subreddit scales with its size. Because our definition of “number of users” includes only those who made at least one comment, each user contributes at least one comment by definition. Therefore, to capture additional activity beyond this minimal baseline in a subreddit, we consider the excess of comments over users, $E = C-U$ where $C$ represents the number of comments, and $U$ the number of users. Figure~\ref{Img_NCom_vs_Nuser} depicts this excess against $U$.

A linear regression of \(\log(C-U)\) against \(\log(U)\) (black line) reveals a strong power-law relationship, with \(r^2 \simeq 0.84\), using a least squares method. Specifically, our data suggest:
\begin{equation}
    \label{Eq:comments_powerlaw}
    C = U + k U^a,
\end{equation}
with \(a \simeq 1.27\), with \(k \simeq 0.33\) and being a proportionality constant.
The first term ($U$) enforces the minimal one-comment-per-user baseline, while the second term indicates super-linear growth in total comments with increasing subreddit size (the `extra activity'). This finding is reminiscent of super-linear scaling observed in other complex systems such as cities and ecosystems~\cite{bettencourt2010urban}.

The regression found in figure \ref{Img_NCom_vs_Nuser} only considered subreddits with more than 10 users, and hereon we will exclude this data from our figures. This choice was made due to the discreteness of the data, which in small regimes generates visual artifacts not representative of the data in the other orders of magnitude. This becomes evident in the case of subreddits of size one, for which inequality in this "individual community" is always zero. Nevertheless, we show in the Supplementary Material that subreddits in the regime \( U \in [1, 10]\) are qualitatively consistent with the main observations we make in this work.

\subsection{User Comments Across Subreddit Sizes}

To explore how this additional activity is distributed among users, we looked at how the distribution of comments-per-user in a subreddit, \(c\), changes with subreddit size. 

\begin{figure}[ht] 
\centering
\includegraphics[width=0.7\textwidth]{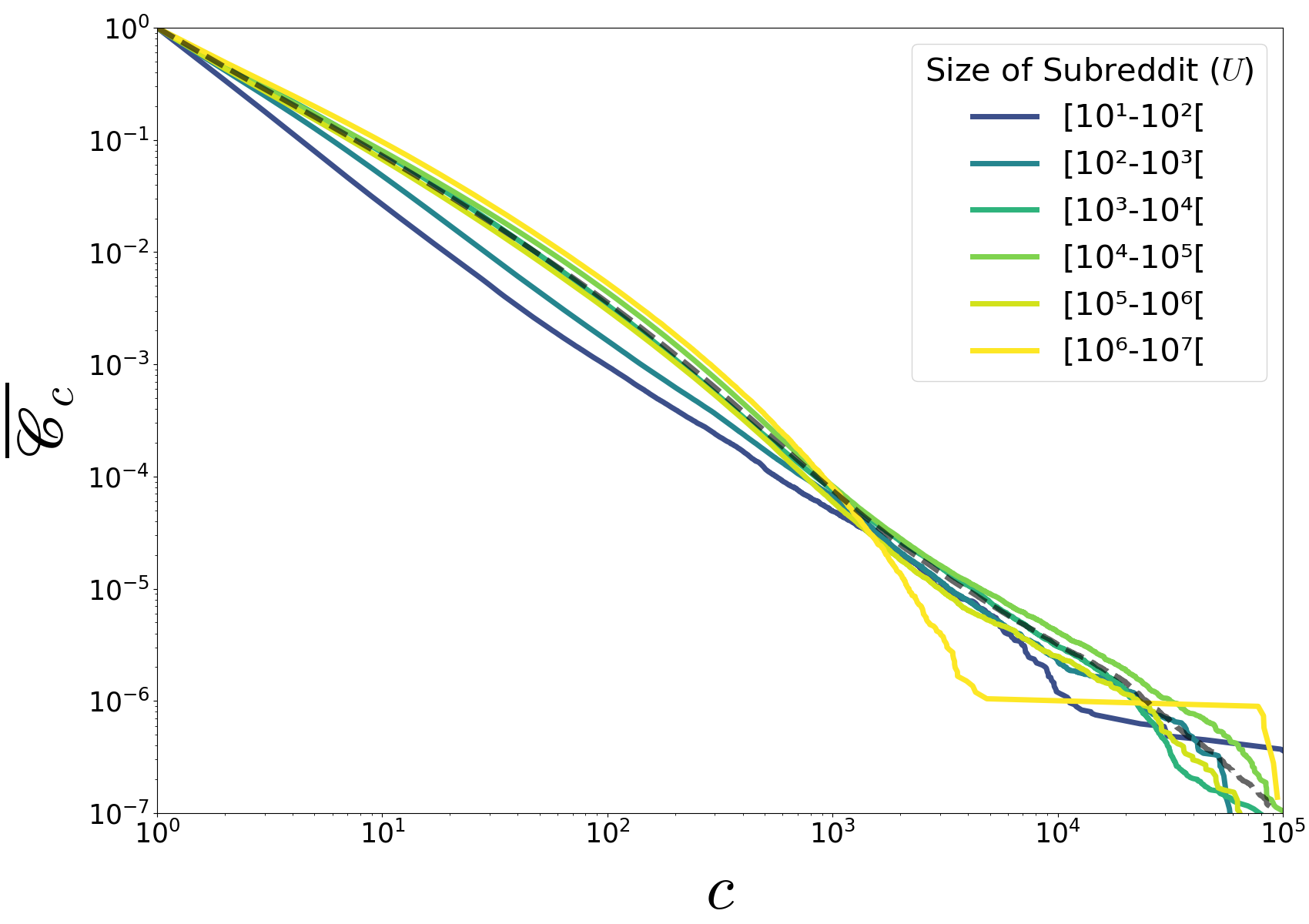}
\caption{The complementary cumulative distribution of the number of comments-per-user in a subreddit, \(c\), averaged over all subreddits in the given size range. Subreddits are binned into logarithmic size ranges. For reference we included the general distribution of comments-per-user in a subreddit (black dashed line), shown in Figure~\ref{Img_Hist_User_Com}. The noise in the end of the distributions is expected in heavy tail distributions, due to the rarity of finding samples for such big values.} 
\label{fig:commentsSizeSubreddit}
\end{figure}

Figure~\ref{fig:commentsSizeSubreddit} shows the complementary cumulative distributions of comments-per-user in a subreddit, \(\mathscr{C}_{c}\). We can see that smaller subreddits (darker curves) display a higher probability of users making very few comments, whereas larger subreddits exhibit wider distributions, indicating user behaviour changes as subreddits grow. Notably, these changes persist under simple normalizations (\eg, dividing by \(U\) or \(\sqrt{U}\), see Supplementary Material), suggesting a complex scaling effect where the distribution itself broadens with subreddit size.

\begin{figure}[ht] 
\centering
\includegraphics[width=0.45\textwidth]{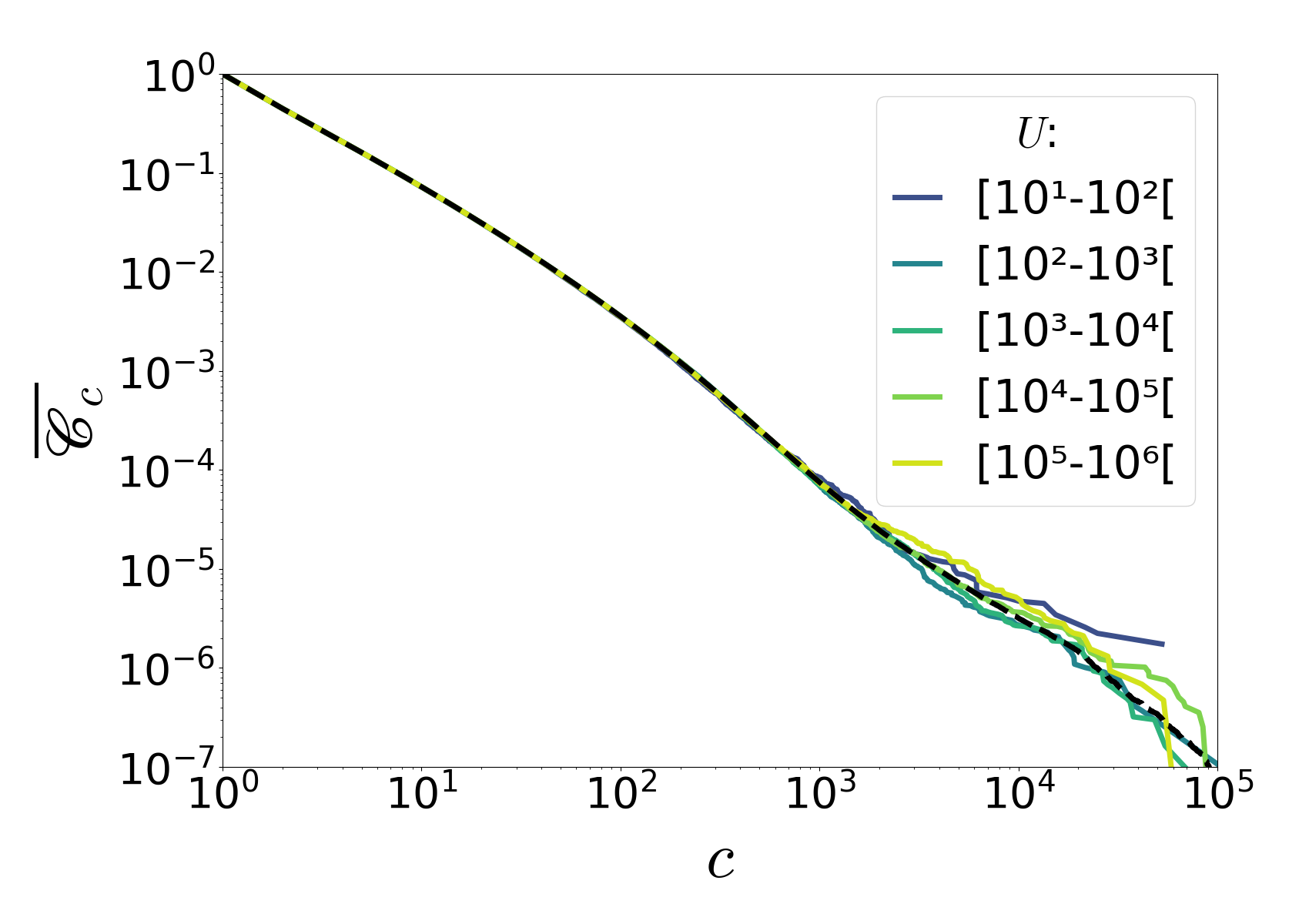}
\includegraphics[width=0.45\textwidth]{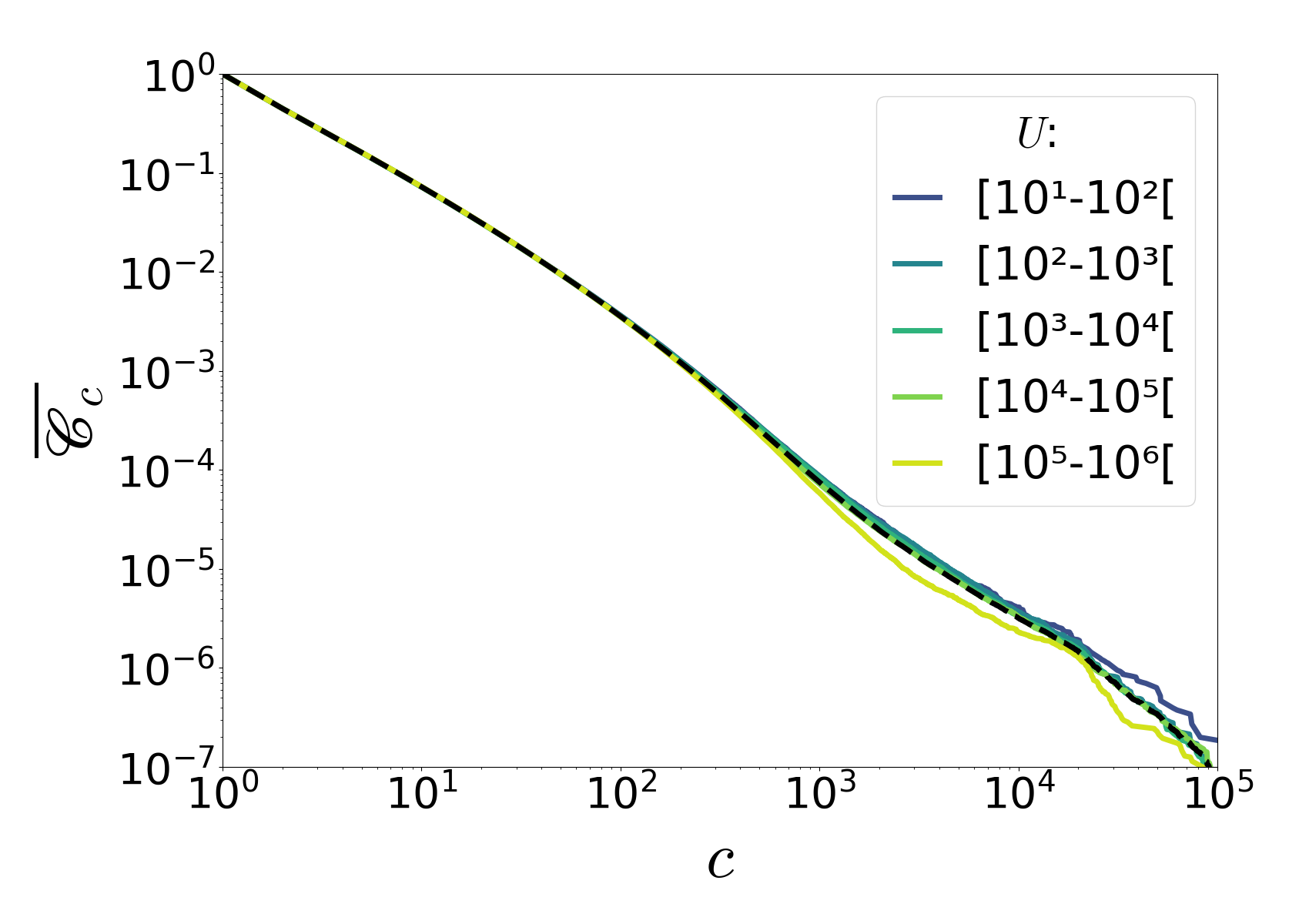}
\caption{(Left Panel) Model 1: The complementary cumulative distribution of comments-per-user made in simulated subreddits, made by sampling from the global comment activity, and averaged over all subreddits of similar size. (Right Panel) Model 2: The complementary cumulative distribution of comments-pet-user made in a simulated subreddit of a certain size, maintaining user profile and total subreddit activity. Both the left and right panels are the equivalents to Figure \ref{fig:commentsSizeSubreddit} for the simulated data, where each line represents the mean of the distributions for all simulated subreddits in a certain size group. 
The black dashed line is the general distribution of comments-per-user of Figure~\ref{Img_Hist_User_Com} added for reference. 
}
\label{fig:models}
\end{figure}

One might suspect that these shifts in distribution could arise from sampling effects: larger subreddits might have a higher chance of including rare, highly active users simply by virtue of sampling more users. 
To test these effects, we developed two null sampling 
models:

\begin{description}
    \item [Model 1: Random sampling from empirical distributions.] 
    We sampled subreddit sizes from the empirical distribution of $U$. Then, for each subreddit, we sampled user comment-counts from the general distribution of comments-per-user in a subreddit (Figure \ref{fig:models}A).

    \item [Model 2: Shuffle subreddits keeping individual user behaviour.]
    We preserved each real user’s number of comments per subreddit, and the total number of users of each subreddit. We then, within each month, randomly assigned the subreddits to each user, while preventing any single user from commenting multiple times in the same subreddit. (Figure \ref{fig:models}B)
\end{description}

In both simulations, the resulting distributions for different subreddit-size groups essentially collapsed back onto the global distribution, diverging only in the extreme tail (where sampling is sparse). This contrasts with the real data (Figure~\ref{fig:commentsSizeSubreddit}), which shows changes in the distributions across the entire range of \(U\), even when averaged over thousands of subreddits. Thus, the systematic broadening we observe cannot be explained by sampling bias alone; it indicates a fundamental shift in user dynamics as subreddits grow, with users not converging to a "mean behaviour". In fact, the exponents characterizing the general distribution of comments-per-user in a month (black dashed line) are associated with a critical regime where the mean and standard deviation diverge.

\subsection{Inequality of User Activity}
 
The Gini coefficient (or Gini index) is a statistical measure used to quantify inequality within a distribution ~\cite{gini1921measurement}. It is widely applied in economics (income disparity), sociology (access to education) and even ecology (biodiversity). The Gini coefficient ranges between 0 (total equality) and 1 (total inequality), and is given by the area between the lorenz curve (ranked cumulative distribution) and the identity line (perfect equality), see Supplementary Material. 

We used the Gini coefficient to quantify how comments become more (or less) concentrated among a subset of users, within a subreddit. The more concentrated the activity is in the most active users, the steeper the initial rise in the ranked cumulative distribution, and hence the larger the Gini coefficient. If activity were evenly distributed among users, the Gini coefficient would be zero; if one user accounted for nearly all comments, it would approach one. Since the standard Gini coefficient has different mathematical limits depending on subreddit size, we normalized each subreddit's Gini coefficient by the maximum possible value for its size \(U\), to enable proper comparison (see Supplementary Materials). Unless stated otherwise, all references to the Gini coefficient refer to this normalized version.

\begin{figure}[htb] 
\centering
\centering
\includegraphics[width=0.45\textwidth]{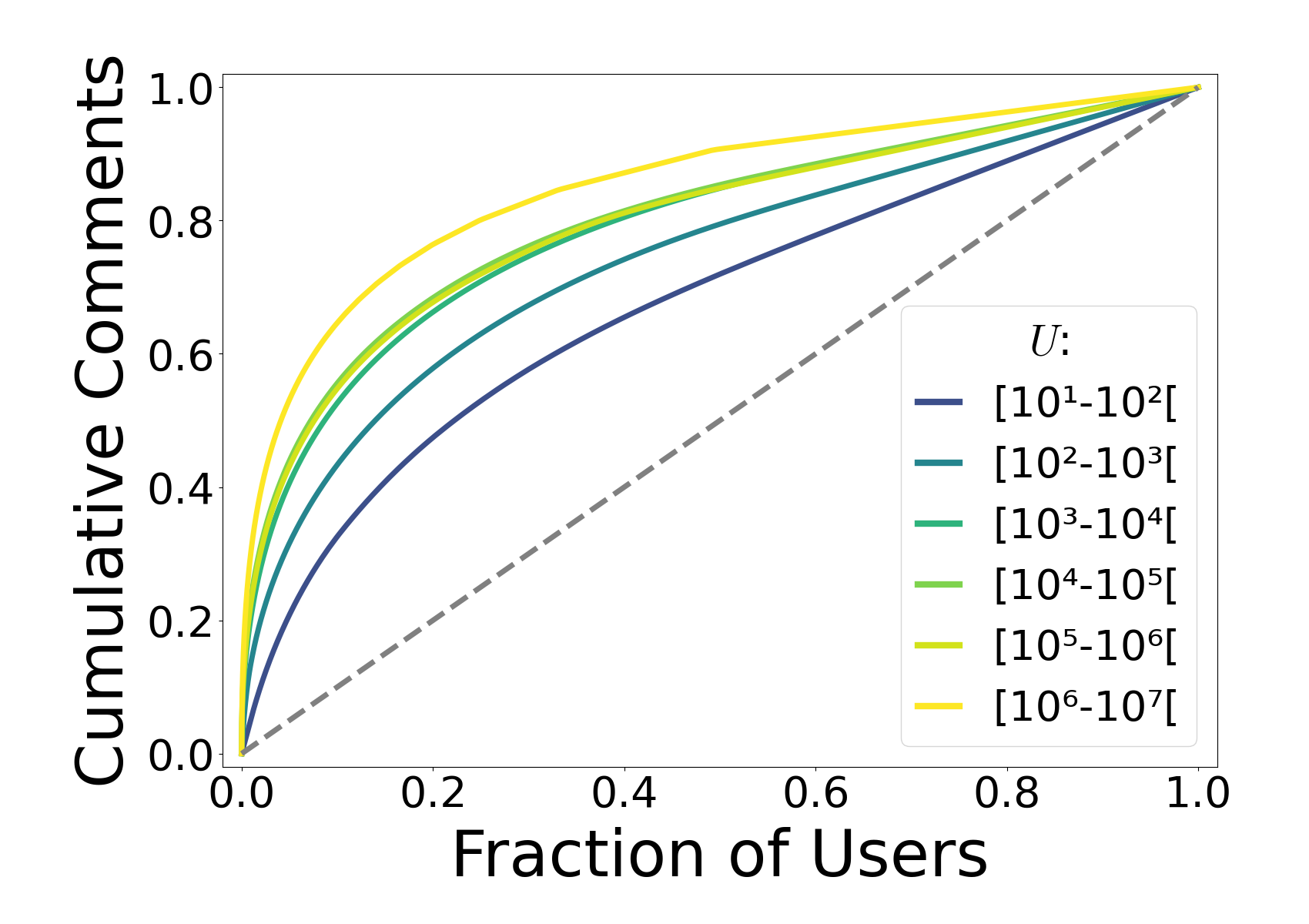}
\includegraphics[width=0.45\textwidth]{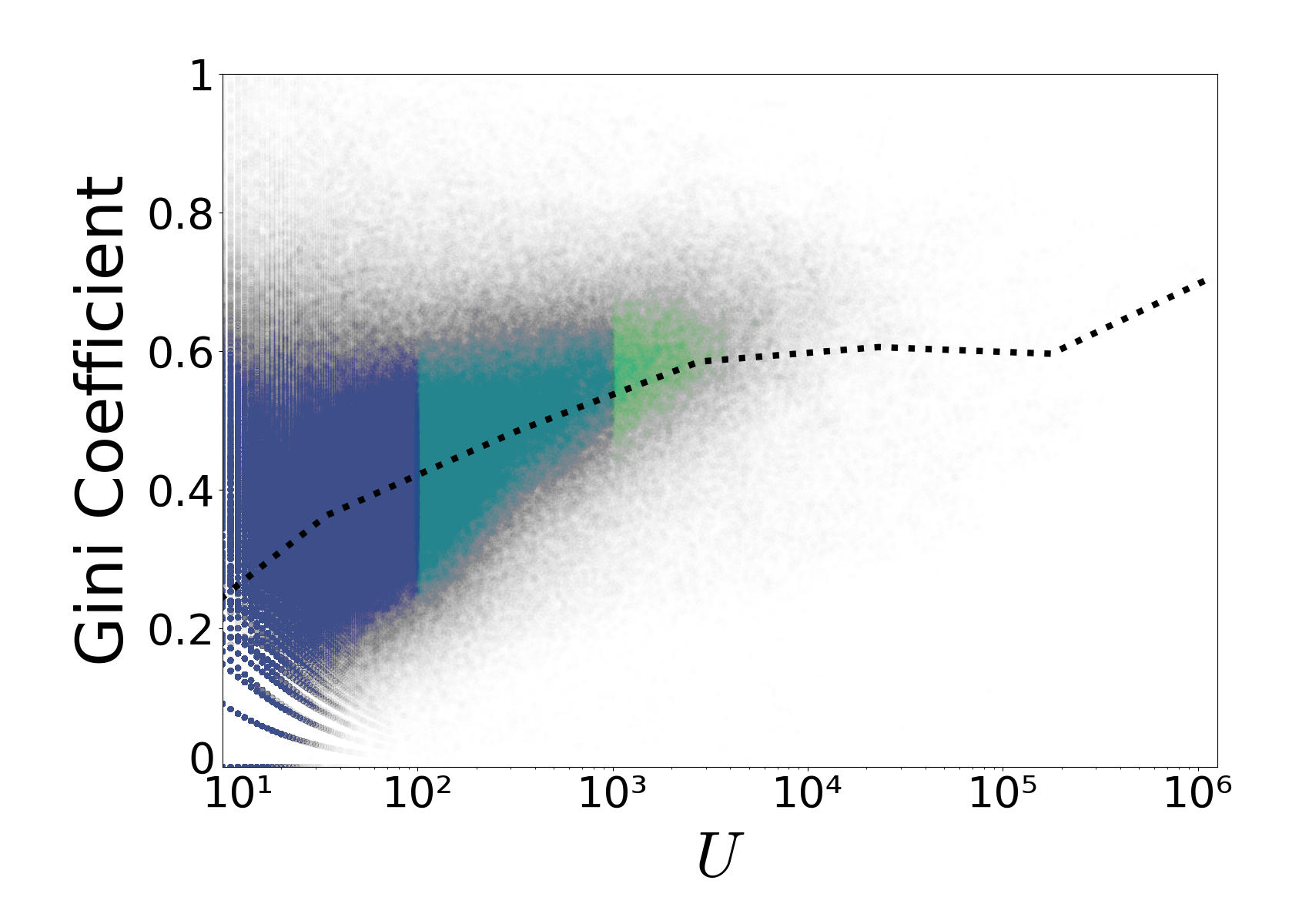}
\caption{(Left) The mean cumulative distribution of comments, for different subreddit sizes. The area below the curves and above the diagonal grey straight line gives the Gini coefficient. Each line represents a group of subreddits of different size, with each point being calculated performing the arithmetic mean of the normalized cumulative comments distribution of each subreddit present in the group. (Right) The Gini coefficient of each subreddit versus the number of users it has. Each dot represents a subreddit. The black dotted line is the mean Gini coefficient of all the subreddits present in a group, sorted by size using a logarithmic scale of ten (as in the other figures), and given by the colour code; the horizontal coordinate is the mean number of users of all the subreddits of such group.} 
\label{fig:Gini}
\end{figure}

Figure~\ref{fig:Gini}A illustrates the average cumulative comment distribution for subreddits of different sizes, while Figure~\ref{fig:Gini}B shows how the Gini coefficient increases with \(U\). In smaller subreddits, 60\% of comments might require 40\% of the users, whereas in larger subreddits, the same 60\% often comes from only 20\% (or fewer) of the users. In other words, as subreddits grow, activity becomes increasingly centralized. However, this increase in the Gini coefficient slows down for very large subreddits, suggesting that there may be practical or structural limits to this inequality.

Several mechanisms might drive this size-dependent rise in inequality. First, larger subreddits tend to produce more posts and interactions overall, possibly amplifying a `rich-get-richer' dynamic among the most active participants. Second, although Reddit’s design and karma system aim to be less algorithmically driven than other platforms, some biases or `visibility loops' could still push attention toward a core of prolific users. Third, bot accounts---whether for moderation or content generation---may disproportionately affect larger communities. Indeed, we identified individual users posting on the order of \(10^7\) comments per month, far exceeding human capabilities.

Ultimately, these findings demonstrate that as subreddits grow, they exhibit super-linear scaling in total activity and an accompanying rise in the concentration of participation. Beyond merely confirming heavy-tailed distributions, our results highlight how community size itself systematically reshapes the underlying patterns of engagement, suggesting that both social and technical factors interact to produce increasing centralization in large online communities.

\section{Discussion, Limitations and Implications}

Our results reveal that as subreddits grow in size, total user activity (as measured by the number of comments) increases in a super-linear manner, indicating that larger communities do not merely accumulate proportional increases in participation. Instead, there is an amplified feedback mechanism that drives engagement beyond a one-to-one relationship between users and comments. This super-linear scaling aligns with patterns observed in other complex systems, such as urban environments, where population size correlates with disproportionately high socioeconomic outputs~\cite{bettencourt2010urban}.

A key insight from this study is the rising inequality of user activity in larger subreddits, wherein an increasingly smaller fraction of highly active users accounts for a large share of comments. Although heavy-tailed distributions are common in online and offline social structures, our results highlight a systematic broadening of the distribution of user activity with subreddit size. Quantifying this effect through the Gini coefficient shows that inequality grows with the number of active users, although eventually plateauing in very large communities. This plateau suggests that structural or platform-related factors may place an upper bound on how concentrated participation can become. Yet, a more interesting alternative could be that there are limits in the inequality in social groups.

These findings must be interpreted in light of several constraints. Mainly, our dataset only allows us to track users who leave comments or posts, excluding those who solely browse or subscribe without interacting; we also lack information on the total number of subscribers to each subreddit over time or on who has viewed or liked specific posts. As a result, our measure of community size is inherently limited to active commenters. Moreover, although Reddit’s openness minimizes certain algorithmic biases found on other platforms, we cannot fully disentangle organic user behaviour from design features (\eg, karma thresholds or default subreddit visibility) that might promote activity among already prominent contributors. We believe the lack of information however does not affect our inequality claims, particularly because non-active users can be considered as an amplification of the inequalities. Yet, it would be interesting to understand what happens to the `quiet' users as the subreddits grow. Are people more likely to be quieter in social interactions if the group is large?

The present study opens multiple avenues for future research. First, an in-depth analysis of user-level behaviours, such as examining users’ entropy or diversity of commenting patterns, could yield deeper insights into how engagement evolves in time or across diverse subreddit themes.
Second, one might explore the influence of geographic or cultural factors. While Reddit does not explicitly provide location data, researchers can sometimes infer regional contexts through language use or community topics.
This line of inquiry may shed light on how local norms or interests shape user participation. Third, analysing explicit user networks (mapping out who interacts with whom) could clarify how certain members become influential hubs, whether through social bonding, common interests, or algorithmic visibility loops. 

From a practical standpoint, understanding the mechanisms that concentrate activity among a small core of users has direct relevance to community moderation and platform governance. Highly active users can be pivotal in creating content and sustaining community interest. Yet, such concentration risks marginalizing less vocal members, thereby reducing diversity of opinion. As platforms like Reddit continue to grow, striking a balance between supporting engaged power users and maintaining equitable participation will be increasingly important.

Lastly, machine learning techniques may further augment our understanding of subreddit dynamics. Predictive models that incorporate historical data on user behaviours, subreddit topics, or even structural changes in the platform could identify emergent trends and automate moderator support or content curation. These algorithms, if designed transparently and ethically, might help pre-empt community imbalances, identify potential moderators, and promote healthier, more diverse engagement. Overall, by situating our findings on super-linear scaling and rising inequality within a broader ecosystem of community dynamics, we underscore both the complexity and the potential for more inclusive design in large-scale online platforms.

\subsection*{Declaration}

For the purpose of open access, the authors have applied a Creative Commons Attribution (CC BY) licence to any accepted manuscript version arising from this submission.

\subsection*{Competing interests}
The authors declare that they have no competing interests.

\subsection*{Funding}
This work was developed within the scope of the project i3N, UIDB/50025/2020 \& UIDP/50025/2020, financed by national funds through the FCT/MEC, and the BioComplex Lab at the University of Exeter.
GM was supported by FCT fellowship 2020.08794.BD .

\subsection*{Author's contributions}
GM and RM developed the original ideas; All authors designed the study; DP, RM and GB supervised the development of the experiments; GM and GB wrote the formalisms; GM collected, curated, and integrated the raw data; GM performed the analysis; All authors analysed the results; All authors wrote the paper; GM and DP prepared the graphics. All authors read, reviewed, and approved the final manuscript.

\bibliographystyle{bmc-mathphys}
\bibliography{bibliography}

\section*{Glossary}

\begin{description}
    \item[\textbf{Active user}] User that made at least one comment in a subreddit during a month.
    \item[\boldmath\(U\)] Total number of active users in a subreddit.
    \item[\boldmath\(C\)] Total number of comments made in a subreddit, during a month.
    \item[\boldmath\(E\)] Excess comments: \(C - U\); the total number of comments made in a subreddit, during one month, minus \(1\) for each user that commented on such subreddit, due to the way we obtain the data.
    \item[\boldmath\(C_{T}\)] Total number of comments made by a user in all subreddits (all of Reddit), during one month.
    \item[\boldmath\(c\)] The number of comments made by a user in a subreddit, during one month.
    \item[\boldmath\(S_{U}\)] The total number of subreddits that have \(U\) active users. 
    \item[\boldmath\(S_{C}\)] The total number of subreddits that have \(C\) comments. 
    \item[\boldmath\(\mathscr{P}_{X}\)] The probability distribution function, \(PDF\), of the variable \(Y\).
    \item[\boldmath\(\mathscr{C}_{X}\)] The complementary cumulative probability distribution, \(CCDF\), of the variable X: \(P_{X}(X \geq x)\).
\end{description}

\end{document}